# Geopolitical Model of Investment Project Implementation

Oleg Malafeyev[1, a)] and Konstantin Farvazov [1, b)] and Olga Zenovich[1, c)]

[1]*St. Petersburg State University, 7/9 Universitetskaya nab., St. Petersburg, 199034, Russia.*

[a)]Corresponding author: malafeyevoa@mail.ru
[b)] konst_far@mail.ru
[c)] olga.zenovich@gmail.com

**Abstract.** Two geopolitical actors implement a geopolitical project that involves transportaion and storage of some commodities. They interact with each other through a transport network. The network consists of several interconnected vertices. Some of the vetrices are trading hubs, storage spaces, production hubs and goods buyers. Actors wish to satify the demand of buyers and recieve the highest possible profit subject to compromise solution principle. A numerical example is given.

## Introduction

With today's levels of industry competition, it proves to be more and more difficult for companies to penetrate new markets. Investors are facing numerous nontrivial tasks that require a formalised approach for finding solutions. A problem of placing company production in the nodes of existing transport network is one of many nontrivial tasks investors face.

In this paper we analyse a model of selecting and implementing an investment project conducted by geopolitical actors. These geopolitical actors interact with each other in a transport network, defined on a plane. The investment project considered in this paper is about building a transport network consisting of trading hubs. Commodities are being delivered to trading hubs from production hubs, located in the vertices of the transport network.

All vertices in the transport network are connected. Thus, there is a path that can be followed from any one node to any other, and this path may not be unique. Several actors interact in this transport network.

Production hubs are located in some vertices of the transport network, and actors are buying commodities there. Moreover, some nodes contain storages, and actors are able to rent these storage spaces for a limited time in order to keep their commodities. This enables actors to decrease storage costs significantly. Moreover, the storage owners charge fees for storing commodities. The fees depend on the amount of commodities to be stored.

Commodities are being delivered from production hubs to storage spaces in batches. Commodities are being delivered to trading hubs from storage spaces in batches as well. Moreover, actors utilize transporting vehicles of different capacities: larger capacity vehicles are used to deliver goods from production hubs to storages; and smaller capacity vehicles are used to deliver goods from storage spaces to trading hubs.

The buyers are located in some nodes of this transport network, and each buyer wishes to obtain a specific amount of goods.

Functions of transport costs for the actors and buyers are defined on the set of arcs. Transport costs are the costs associated with moving on the arc. Transport cost functions for actors differ depending on the vehicle used: larger capacity vehicles incur bigger costs than smaller capacity ones.

Cumulative costs for the buyer consist of prices to be paid to get a required amount of goods and costs associated with transport. Buyers decide which trading hub to choose as they tend to minimize their cumulative costs.

A set of nodes of possible locations of trading hubs in the network is also defined. Each actor can place his/her trading hub in one of the nodes inside this set. Thus, each actor chooses production hub and storage spaces, so that his/her cumulative costs for producing, storing and delivering goods is minimal.

Actors also wish to place trading hubs in the most lucrative way in terms of profit maximization from selling the goods.

Thus, actors' investment project is deemed to be successful if they solve the following problem: a finite number of trading hubs needs to be placed at the nodes of the transport network with given locations of production hubs, buyers and storages and in accordance with optimality principle. In terms of optimality principle we consider a compromise solution.

Both Russian and foreign literature have been used in analyzing the above problem.

Detailed information on game theory and its applications is presented in [1]. Work [2] details information on network flows and also describes algorithms for finding multicriteria solutions for various optimality principles. Problems from matrix games and linear programming together with static and dynamic production models are considered in [3]. Information on graph theory, description of shortest path and algorithms of finding it are discussed in [4]. Paper [5] describes Hotelling's linear city model which is widely used to analyze competition. The basis of economic theory is presented in [6]. Some other relevant models are studied in [7-99]

## Formulation of the Problem of a Finite Number of Trading Hubs Location at the Transport Network Nodes According to Compromise Solution Principle

We define transport network $(N,k)$, where $N$ – is a finite number of nodes, $k$ – is a congestion function that assigns a number $k(x,y) \geq 0$ to a network's arc $(x,y)$, $x,y \in N$. There exist $m$ geopolitical actors who work on an investment project about constructing a trading network consisting of $m$ trading hubs.

$S$ number of buyers are located at the vertices belonging to $P=(p_1,\ldots,p_s)$ set. Each buyer wishes to obtain a specific amount of commodities. The amounts of goods to be bought by buyers are defined by vector $V=(v_1,\ldots,v_s)$.

For each arc $(x,y)$ we define a function of transportation costs for actors: $\acute{C}_a^1(x,y) \geq 0$ for higher capacity transport and $\acute{C}_a^2(x,y) \geq 0$ for lower capacity transport, $\acute{C}_a^1(x,y) \geq \acute{C}_a^2(x,y) \forall (x,y)$. A function of transportation costs for buyers $\acute{C}_b(x,y) \geq 0$ is also defined.

$H$ number of production hubs are located in some nodes $D=(d_1,\ldots,d_h)$ of the network, and this is where actors are buying their commodities from. The costs of buying goods at different production hubs are defined by vector $L=(l_1,\ldots,l_h)$.

$R$ storages are located in some nodes $K=(k_1,\ldots,k_r)$ of the network, and this is where actors are able to store bought goods. The owners of storage spaces are charging for keeping goods, and these charges depend on the amount of commodities to be stored. The costs for storing goods at different storage spaces are defined by vector $S=(s_1,\ldots,s_r)$. From production hubs to storage spaces the commodities are delivered in fixed batches of size $Q^1$ by transport with higher capacity.

The actors have a possibility to place their trading hubs in some free nodes $G=(g_1,\ldots,g_c)$, $c \geq m$, of the network. Thus, we have $m$ trading hubs inside set $G$, and these hubs belong to set $T=(t_1,\ldots,t_m)$, $T \in G$, and each of these trading hubs is receiving commodities in fixed batches of size $Q^2$ by transport with lower capacity coming from storage spaces.

Actor selects production hub and storage space so that his/her cumulative costs for buying, storing and delivering commodities are minimal.

The unit cost $P_i$ of commodity at the trading hub $i$, $i=\overline{1,m}$ is comprised of the following parts: the sum the commodity was bought for by the actor at production hub, the cost of storing it, the cost of transporting it and some margin, added by the actor:

$$P_i = l_i + s_i + \frac{C_i^1}{Q^1} + \frac{C_i^2}{Q^2} + w_i$$

where: $l_i$ – is cost of a commodity unit paid by the actor at the production hub; $s_i$ – storage rent; $C_i^1$ – transport costs from the production hub to the storage space; $C_i^2$ – transport costs from the storage space to the trading hub; $w_i$ – is an additional margin.

Cumulative costs $U_j$ for a client $j$, $j=\overline{1,s}$ are comprised of a price to pay for a required amount of commodities at the trading hub $i$, $i=\overline{1,m}$ and own cumulative transport costs spent on commuting to this trading hub:

$$U_j = v_j P_i + C_i^3$$

where: $v_j$ – number of commodity units bought by a buyer $j$; $P_i$ – commodity unit price at the trading hub $i$; $C_i^3$ – transport costs for commuting to the trading hub $i$.

Each buyer $j$ tends to minimize his/her cumulative costs, i.e. to find $\min_i (v_j P_i + C_i^3)$. Hence it is importatnt for a buyer to choose a specific trading hub.

Each actor wishes to place his/her trading hub in a most profitable way, so that the revenue from selling commodities is maximized. For the investment project to be successful, the actors need to solve a following problem – how to place a finite number of trading hubs at the nodes of the transport network with given locations of production hubs, buyers and storages and in accordance with compromise solution principle.

## Solution to the problem of placing a finite number of trading hubs at the transport network nodes according to compromise solution principle

The process of finding solution to this problem can be divided into several parts:
1. First part is to find paths which provide minimal transport costs from production hubs to storages, and from storages to trading hubs for the actors; and paths which provide minimal transport costs for the buyers to commute to trading hubs.
2. Second part is to build a matrix of revenues generated by the trading hubs depending on their location.
3. Last part is to find a compromise solution.

## Floyd-Warshall algorithm for finding shortest paths between all vertices in a weighted graph

Consider a transport network. At each arc of the transport graph a transport cost function is defined. To solve the above problem we need to find such paths between graph's vertices, so that transport costs are minimal. One way to find such paths would be to use Floyd-Warshall algorithm.

Consider an edge-weighted graph $G = (V, E)$. It is required to find the shortest paths between all pairs of vertices. Assume that there are no cycles whose edges sum to a negative value.

Build a $V \times V$ matrix $D^0$ with elements defined according as follows:

$$\begin{cases} d_{ii}^0 = 0 \\ d_{ij}^0 = weight(v_i, v_j), i \neq j, if\ edge(v_i, v_j) \exists \\ d_{ij}^0 = \infty, i \neq j, if\ edge(v_i, v_j) \nexists \end{cases}$$

Consider then $m := 0$ and build matrix $D^{m+1}$ by using $D^m$:

$$\begin{cases} d_{ii}^{m+1} = 0 \\ d_{ij}^{m+1} = \min\{d_{ij}^m, d_{i,m+1}^m + d_{m+1,j}^m\}, i \neq j \end{cases} \quad (1)$$

If for any $i$, $d_{i\mathfrak{z}}^m + d_{mi}^m < 0$, then a negative cycle going through the vertex $v_i$ exists.

Consider then $m := m+1$. Repeat step (1) until $m < V$. As a result, when $m = V$, we have matrix $D^V$, whose elements are the lengths of the shortest paths between corresponding vertices.

## Algorithm for finding a compromise solution

In order to find a compromise solution in a problem of placing trading hubs, one needs to know payoff function of each actor. In the problem described above, the actor's payoff function corresponds to the price paid by a buyer to obtain commodity.

Consider a matrix of payoffs in different trading hubs depending on hubs' location. The rows in the payoff matrix $\Gamma$ correspond to the trading hubs, and columns correspond to possible game situations:

$$\Gamma = (\alpha_{m,q})$$

where: $m$ – is a number of trading hubs; $q$ – a number of situations inside the game.

Build a perfect vector consisting of maximum values of trading hubs' payoffs:

$$M = \begin{pmatrix} M_1 \\ \ldots \\ M_m \end{pmatrix}, \text{where } M_i = \max_q (\alpha_{m,q}), i = \overline{1, m}$$

In order to build a residuals matrix, one needs to compute deviations of payoffs from maximum payoffs for each trading hub:

$$\Gamma_M = (M - \alpha_{m,q}) = (\beta_{m,q})$$

Now in each situation in the residuals matrix sort the values in an ascending order, so that first row contains smallest residuals, and last row contains largest residuals. Thus, last row contains maximal residuals $\max_m (\beta_{m,q})$.

Finally, find minimal value across maximal residuals $\min_q \max_m (\beta_{m,q})$.

The derived situation is a compromise solution.

If multiple minimal values exist in the last row, then one should look for minimal value in the last but one row, and so on. This means that several situations with compromise solution exist.

# Numerical example of placing 2 trading hubs in a transport network composed of 30 vertices and 50 edges with a given location of 2 production hubs, 2 storages and 4 buyers

Consider a network with $N$ with 30 vertices $x_0, \ldots, x_{29}$ and 50 edge, shown on Fig. 1. Each edge has values of transport cost function for both actors and buyers assigned to them (for high capacity and low capacity transport). There exist 2 geopolitical actors in the network, and they wish to place their trading hub inside this network.

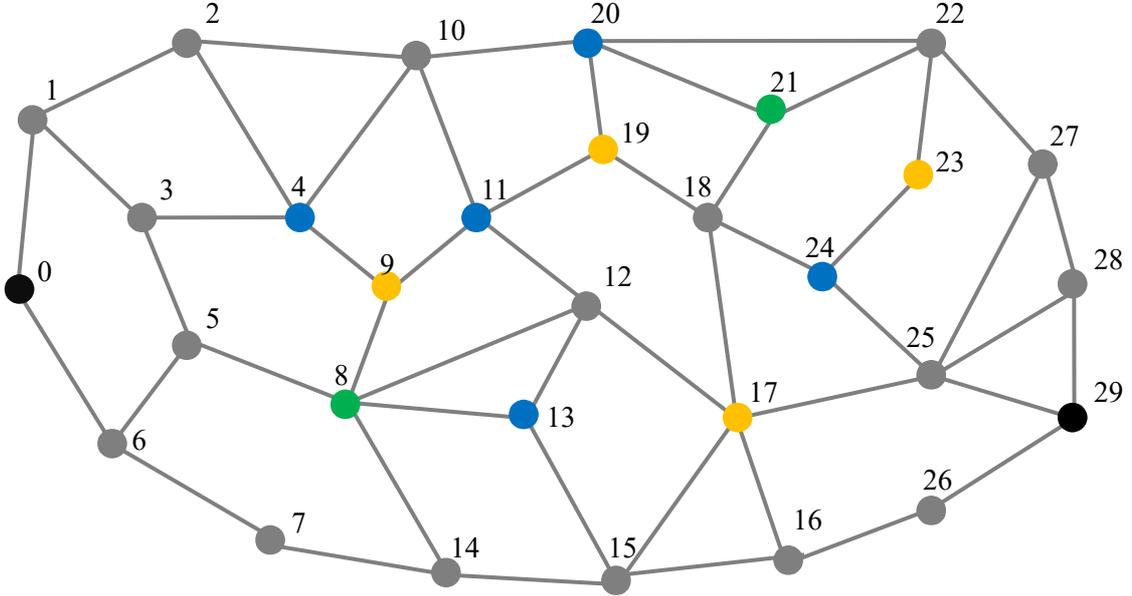

**FIGURE 1.** Transport network.

Production hubs are located in the nodes $x_0$ and $x_{29}$, and this is where actors obtain their commodities from (the vertices are shown in black). The cost of one commodity unit at these production hubs is $l_1=4$ and $l_2=2$ respectively.

The buyers are located at the vertices $x_4$, $x_{13}$, $x_{20}$, $x_{24}$, and they wish to buy commodities (the vertices are shown in blue). Each buyer wants to buy 5 units of commodity, i.e. $v_1=v_2=v_3=v_4=5$.

The storages are located in the vertices $x_8$ and $x_{21}$, and this is where actors can store their goods (the vertices are shown in green). The cost for storing 1 unit of commodity is $s_1=1,5$ and $s_2=1$ respectively. The actors deliver goods from production hubs to storage spaces in batches of $Q^1=10$.

The actors have a possibility to place their trading hubs in 4 vertices $x_9$, $x_{17}$, $x_{19}$ and $x_{23}$ (the vertices are shown in yellow). The goods are delivered from storages to trading hubs in batches of $Q^2=5$.

The margin inside the price of one commodity unit accounts for 50% of the sum the commodity was bought for by the actor at production hub, the cost of storing it and the cost of transporting it:

$$w_i = \frac{1}{2}\left(l_i + \frac{C_i^1}{Q^1} + \frac{C_i^2}{Q^2} + s_i\right)$$

Transport costs for actor for using higher capacity vehicles are $\acute{C}_a^1(x,y)$ and for using lower capacity vehicles $\acute{C}_a^2(x,y)$. Transport costs for buyers are $\acute{C}_b(x,y)$. All costs are shown in table 1.

**TABLE 1.** Transport costs.

| $(x,y)$ | $\acute{C}_a^1($ | $\acute{C}_a^2($ | $\acute{C}_b(x$ |
|---|---|---|---|
| (0,6) | 6 | 4 | 7 |
| (5,6) | 12 | 8 | 10 |

| Edge | | | |
|---|---|---|---|
| (0,1) | 3 | 2 | 8 |
| (1,3) | 10 | 9 | 5 |
| (1,2) | 13 | 6 | 13 |
| (3,4) | 19 | 12 | 7 |
| (3,5) | 11 | 7 | 8 |
| (5,8) | 6 | 2 | 2 |
| (6,7) | 21 | 16 | 7 |
| (7,14) | 10 | 6 | 9 |
| (8,14) | 19 | 15 | 9 |
| (8,13) | 16 | 9 | 5 |
| (8,9) | 10 | 7 | 13 |
| (4,9) | 9 | 5 | 9 |
| (4,10) | 17 | 13 | 5 |
| (2,4) | 14 | 12 | 14 |
| (2,10) | 15 | 11 | 9 |
| (10,11) | 6 | 5 | 7 |
| (11,19) | 12 | 7 | 3 |
| (10,20) | 15 | 10 | 11 |
| (19,20) | 16 | 12 | 5 |
| (20,21) | 10 | 6 | 8 |
| (20,22) | 23 | 14 | 15 |
| (21,22) | 12 | 10 | 14 |
| (18,21) | 8 | 6 | 6 |
| (18,19) | 9 | 5 | 2 |
| (11,12) | 24 | 13 | 9 |
| (9,11) | 15 | 10 | 3 |
| (8,12) | 13 | 8 | 8 |
| (12,13) | 13 | 7 | 8 |
| (12,17) | 17 | 10 | 14 |
| (17,18) | 5 | 4 | 5 |
| (18,24) | 12 | 10 | 2 |
| (23,24) | 19 | 10 | 13 |
| (22,23) | 16 | 11 | 8 |
| (22,27) | 9 | 6 | 9 |
| (27,28) | 21 | 15 | 14 |
| (25,27) | 6 | 3 | 7 |
| (24,25) | 16 | 13 | 11 |
| (17,25) | 7 | 5 | 9 |
| (13,15) | 12 | 6 | 3 |
| (15,17) | 14 | 8 | 6 |
| (14,15) | 15 | 10 | 6 |
| (15,16) | 9 | 5 | 9 |
| (16,17) | 20 | 15 | 4 |
| (16,26) | 14 | 11 | 14 |
| (26,29) | 19 | 17 | 9 |
| (25,29) | 9 | 7 | 3 |
| (25,28) | 14 | 8 | 4 |
| (28,29) | 16 | 13 | 10 |

First find paths for actors with minimal transport costs from production hubs to storages. As higher capacity transport is used, matrix of weights of graph's edges $D_1^0$ is composed of transport costs $\acute{C}_a^1(x,y)$. By applying Floyd-Warshall algorithm, we obtain matrix composed of paths with the smallest transport costs for the actors between all pairs of vertices when higher capacity transport is used.

Thus, the weights of paths with minimal transport costs from production hubs to storages are calculated and shown in table 2.

TABLE 2. Weights of paths with minimal transport costs.

|  | $x_8$ | $x_{21}$ |
|---|---|---|
| $x_0$ | 24 | 56 |
| $x_{29}$ | 46 | 29 |

As the size of the batch $Q^1=10$, the minimal transport costs on the way from production hubs to storages are as shown in table 3.

TABLE 3. Minimal transport costs from production hubs to storages.

|  | $x_8$ | $x_{21}$ |
|---|---|---|
| $x_0$ | 2,4 | 5,6 |
| $x_{29}$ | 4,6 | 2,9 |

Now find paths for the actors with minimal transport costs from storages to trading hubs. In this case commodities are delivered by smaller capacity transport. Matrix of weights of graph's edges $D_2^0$ is composed of transport costs $\acute{C}_a^2(x,y)$. By applying Floyd-Warshall algorithm again, we obtain matrix composed of paths with the smallest costs for the actors between all pairs of vertices when smaller capacity transport is used.

Thus, the weights of paths with minimal transport costs from storages to the possible locations of trading hubs can be found. These weights are listed in table 4.

TABLE 4. Weights of paths with minimal transport costs from storages to the trading hubs.

|  | $x_9$ | $x_{17}$ | $x_{19}$ | $x_{23}$ |
|---|---|---|---|---|
| $x_8$ | 7 | 18 | 24 | 42 |
| $x_{21}$ | 28 | 10 | 11 | 21 |

As the size of the batch $Q^2=5$, the minimal transport costs on the way from storages to the trading hubs are as shown in table 5.

TABLE 5. Minimal transport costs on the way from storages to the trading hubs.

|  | $x_9$ | $x_{17}$ | $x_{19}$ | $x_{23}$ |
|---|---|---|---|---|
| $x_8$ | 1,4 | 3,6 | 4,8 | 8,4 |
| $x_{21}$ | 5,6 | 2 | 2,2 | 4,2 |

Thus, the cumulative transport costs for the actor to deliver 1 unit of commodity to the possible locations of the trading hubs (depending on the location choice for production hub and storage space) are comprised in table 6.

TABLE 6. Transport costs to deliver 1 unit of commodity.

|  | $x_9$ | $x_{17}$ | $x_{19}$ | $x_{23}$ |
|---|---|---|---|---|
| $x_0 \to x_8$ | 3,8 | 6 | 7,2 | 10,8 |
| $x_0 \to x_{21}$ | 7 | 9,2 | 10,4 | 14 |
| $x_{29} \to x_8$ | 6 | 8,2 | 9,4 | 13 |
| $x_{29} \to x_{21}$ | 8,5 | 4,9 | 5,1 | 7,1 |

So, by considering the cost of 1 unit of commodity in the production hubs ( $l_1=4$ in the production hub $x_0$ and $l_2=2$ in the production hub $x_{29}$ ); cumulative transport costs; storage costs ( $s_1=1,5$ at the storage $x_8$ and $s_2=1$ at the storage $x_{21}$ ); and a margin of 50%, we obtain the cost of 1 unit of commodity in all possible locations (vertices) of the trading hub are as listed in table 7.

TABLE 7. Total costs to deliver 1 unit of commodity.

|  | $x_9$ | $x_{17}$ | $x_{19}$ | $x_{23}$ |
|---|---|---|---|---|
| $x_0 \to x_8$ | 13,95 | 17,25 | 19,05 | 24,45 |

| $x_0 \to x_{21}$ | 18    | 21,3  | 23,1  | 28,5  |
|------------------|-------|-------|-------|-------|
| $x_{29} \to x_8$ | 14,25 | 17,55 | 19,35 | 24,75 |
| $x_{29} \to x_{21}$ | 17,25 | 11,85 | 12,15 | 15,15 |

Considering the fact, that actor selects production hub and storage space locations, so that cumulative costs for buying, storing and delivering goods are minimal, we can find the final cost per 1 commodity unit in all possible locations of placing the trading hubs. The results are shown in table 8.

TABLE 8. Final cost per 1 unit of commodity.

| $x_9$ | $x_{17}$ | $x_{19}$ | $x_{23}$ |
|-------|----------|----------|----------|
| 13,95 | 11,85    | 12,15    | 15,15    |

Now, considering that buyers are willing to buy $v_1=v_2=v_3=v_4=5$ commodity units, we can calculate the cost of buying required amount of goods for each buyer depending on the location of trading hub. These costs comprise table 9.

TABLE 9. Final cost per 1 unit of commodity.

|          | $x_9$ | $x_{17}$ | $x_{19}$ | $x_{23}$ |
|----------|-------|----------|----------|----------|
| $x_4$    | 69,75 | 59,25    | 60,75    | 75,75    |
| $x_{13}$ | 69,75 | 59,25    | 60,75    | 75,75    |
| $x_{20}$ | 69,75 | 59,25    | 60,75    | 75,75    |
| $x_{24}$ | 69,75 | 59,25    | 60,75    | 75,75    |

Now, we need to find paths from buyer location vertices to the trading hubs with minimal costs. Matrix of weights of graph's edges $D_3^0$ is composed of transport costs $\acute{C}_b(x,y)$. By applying Floyd-Warshall algorithm, we obtain matrix composed of paths with the smallest transport costs between all pairs of vertices for the buyers.

Thus, the weights of paths with minimal transport costs from buyers location nodes to the possible locations of trading hubs are shown in table 10.

TABLE 10. Weights of paths with minimal transport costs from buyers to trading hubs.

|          | $x_9$ | $x_{17}$ | $x_{19}$ | $x_{23}$ |
|----------|-------|----------|----------|----------|
| $x_4$    | 9     | 22       | 15       | 32       |
| $x_{13}$ | 18    | 9        | 16       | 29       |
| $x_{20}$ | 11    | 12       | 5        | 22       |
| $x_{24}$ | 10    | 7        | 4        | 13       |

By summing up transport costs to reach trading hub and costs of buying required amount of goods, we obtain cumulative costs for the buyer, which depend on the choice of the trading hub. The costs are listed in table 11.

TABLE 11. Cumulative costs for the buyer.

|          | $x_9$ | $x_{17}$ | $x_{19}$ | $x_{23}$ |
|----------|-------|----------|----------|----------|
| $x_4$    | 78,75 | 81,25    | 75,75    | 107,75   |
| $x_{13}$ | 87,75 | 68,25    | 76,75    | 104,75   |
| $x_{20}$ | 80,75 | 71,25    | 65,75    | 97,75    |
| $x_{24}$ | 79,75 | 66,25    | 64,75    | 88,75    |

Now, considering the fact, that each buyer chooses location of the trading hub to minimize his/her cumulative costs, we can build matrix of trading hubs' payoffs depending on their location. In total, 6 possible profiles exist. Profiles and corresponding payoffs are listed in table 12.

TABLE 12. Trading hubs' payoffs.

|   | (9,17)  | (9,19) | (9,23) | (17,19) | (17,23) | (19,23) |
|---|---------|--------|--------|---------|---------|---------|
| 1 | 69,75   | 0      | 279    | 59,25   | 237     | 243     |
| 2 | 177,75  | 243    | 0      | 182,25  | 0       | 0       |

This matrix is a payoffs matrix $\Gamma$. Trading hubs are defined in the matrix rows, and possible situations are defined in the matrix columns:

$$\Gamma = (\alpha_{m,q}); m=1,2; q=1,\ldots,6$$

The perfect vector M is as follows:

$$M = \begin{pmatrix} M_1 \\ M_2 \end{pmatrix} = \begin{pmatrix} \max_q \alpha_{1,q} \\ \max_q \alpha_{2,q} \end{pmatrix} = \begin{pmatrix} 279 \\ 243 \end{pmatrix}$$

Table 13 containes the residuals matrix $\Gamma_M = (\beta_{m,q}) = (M - \alpha_{m,q})$.

**TABLE 13.** The residuals matrix $\Gamma_M$.

|   | (9,17) | (9,19) | (9,23) | (17,19) | (17,23) | (19,23) |
|---|--------|--------|--------|---------|---------|---------|
| 1 | 209,25 | 279    | 0      | 219,75  | 42      | 36      |
| 2 | 65,25  | 0      | 243    | 60,75   | 243     | 243     |

Now in each situation in the residuals matrix sort the values in an ascending order, so that first row contains smallest residuals, and last row contains largest residuals. Then we the follwoing matrix contained in table 14.

**TABLE 14.** Sorted residuals matrix $\Gamma_M$.

|   | (9,17) | (9,19) | (9,23) | (17,19) | (17,23) | (19,23) |
|---|--------|--------|--------|---------|---------|---------|
| 1 | 65,25  | 0      | 0      | 60,75   | 42      | 36      |
| 2 | 209,25 | 279    | 243    | 219,75  | 243     | 243     |

Finally, find minimal value across maximal residuals in the last row

$$\min_q \max_m (\beta_{m,q}) = 209,25$$

Situation (9,17) corresponds to the obtained level. This situation is a compromise solution. Thus, according to the compromise solution principle, trading hubs should be placed in the vertices $x_9$ and $x_{17}$. The actors' payoffs will be: (69,75; 177,75).

## Conclusion

Interaction between two geopolitical actors is studied in this paper. The agents try to maximize their payoffs by placing trading hubs in vertices of the transport network. Floyd-Warshall algorithm for finding shortest paths between all vertices of a weighted graph and the compromise solution principle is used to solve the optimization problem associated with the model. A numerical example of the transport network composed of 30 vertices and 50 edges is considered; positions of 2 production hubs, 2 storages, and 4 buyers is given. The optimal location of trading hubs in the transport network subejct to the compromise solution principle is found.